\documentclass[manuscript]{aastex}
\usepackage{lineno}
\usepackage{bm}
\usepackage{color}
\usepackage{esint}
\usepackage{appendix}
\usepackage{multirow}
\usepackage{makecell}
\usepackage{threeparttable}
\usepackage{url}
\usepackage{ragged2e}
\usepackage{longtable}
\usepackage{rotating}
\usepackage{footnote}
\usepackage{diagbox}

\begin{document}
\title{Acceleration and Expansion of a Coronal Mass Ejection in the High Corona: Role of Magnetic Reconnection}
\author{Bin Zhuang\altaffilmark{1}, No\'{e} Lugaz\altaffilmark{1}, Manuela Temmer\altaffilmark{2}, Tingyu Gou\altaffilmark{3}, and Nada Al-Haddad\altaffilmark{1}}

\affil{$^1$Institute for the Study of Earth, Oceans, and Space, University of New Hampshire, Durham, NH, USA; \url{bin.zhuang@unh.edu}, \url{noe.lugaz@unh.edu} \\
	$^2$Institute of Physics, University of Graz, A-8010 Graz, Austria \\
	$^3$CAS Key Laboratory of Geospace Environment, Department of Geophysics and Planetary Sciences, University of Science and Technology of China, Hefei 230026, China \\ } 

\begin{abstract}
The important role played by magnetic reconnection in the early acceleration of coronal mass ejections (CMEs) has been widely discussed. However, as CMEs may have expansion speeds comparable to their propagation speeds in the corona, it is not clear whether and how reconnection contributes to the true acceleration and expansion separately. To address this question, we analyze the dynamics of a moderately fast CME on 2013 February 27, associated with a continuous acceleration of its front into the high corona, even though its speed had reached $\sim$700~km~s$^{-1}$ and larger than the solar wind speed. The apparent CME acceleration is found to be due to the CME expansion in the radial direction. The CME true acceleration, i.e., the acceleration of its center, is then estimated by taking into account the expected deceleration caused by the solar wind drag force acting on a fast CME. It is found that the true acceleration and the radial expansion have similar magnitudes. We find that magnetic reconnection occurs after the CME eruption and continues during the CME propagation in the high corona, which contributes to the CME dynamic evolution. Comparison between the apparent acceleration related to the expansion and the true acceleration that compensates the drag shows that, for this case, magnetic reconnection contributes almost equally to the CME expansion and to the CME acceleration. The consequences of these measurements for the evolution of CMEs as they transit from the corona to the heliosphere are discussed.
\end{abstract}

\clearpage

\section{Introduction}\label{intro}
\justifying
Coronal mass ejections (CMEs) are large-scale eruptions of coronal plasma and magnetic fluxes ejected into the heliosphere. The speed of CMEs ranges from $\sim$100--3500~km~s$^{-1}$ \citep{2019SSRv..215...39L}, and  fast CMEs (e.g., speed $>$800~km~s$^{-1}$) are of great interest for their role in causing extreme space weather events \citep[e.g.,][]{kilpua2019,2018E&PP....2..112Z,2020ApJ...901...45Z}. Typically, CMEs are found to undergo three phases of dynamic evolution: (1) an initiation phase with a slow rise, (2) an impulsive phase with a fast acceleration, and (3) a propagation phase with a constant or slowly decreasing speed \citep{2001ApJ...559..452Z,2004ApJ...604..420Z,2008ApJ...673L..95T,2010ApJ...712.1410T}. During these three phases, the dominant force acting on CMEs is different \citep[e.g.,][]{2004A&A...423..717V,2015SoPh..290..903M}, i.e., the Lorentz force competing with the solar wind drag force (solar gravity can be neglected), as further explained below.

In the low corona where magnetic pressure is dominant over thermal pressure, the Lorentz force accounts for the CME initiation and initial acceleration \citep[e.g.,][]{2000JGR...10523153F}. The acceleration usually takes place within a few to tens of minutes impulsively with magnitudes varying from $\sim$200 to $\sim$7000~m~s$^{-2}$ \citep[e.g.,][]{2001ApJ...559..452Z,2004ApJ...604..420Z,2007SoPh..241...85V,2008ApJ...673L..95T}, and leads to fast CMEs associated with solar flares. Two temporal correlations, i.e., (1) between the solar flare soft X-ray flux and CME speed profile \citep[e.g.,][]{2001ApJ...559..452Z,2004ApJ...604..420Z,2004SoPh..225..355V}, and (2) between the flare hard X-ray flux and the acceleration profile of fast CMEs \citep{2008ApJ...673L..95T,2010ApJ...712.1410T,2018ApJ...868..107V,2020ApJ...897L..36G,2020ApJ...893..141Z}, have revealed a close flare-CME relationship. Flares and CMEs are coupled through the action of magnetic reconnection \citep[e.g.,][]{2020ApJ...897L..36G}. On the one hand, magnetic reconnection can result in a very large amount of energy release, which contributes to the plasma heating and the acceleration of solar energetic particles in flares. On the other hand, reconnection plays a significant role in the CME acceleration, which is achieved via reducing the tension force owing to the overlying magnetic loops and providing additional magnetic (especially poloidal) flux to the original CME structure \citep{2000JGR...105.2375L,2004ApJ...602..422L,2008AnGeo..26.3089V}.

At larger distances where coronal magnetic field decreases significantly and the solar wind has reached a faster speed, CMEs primarily experience an aerodynamic drag force due to interaction with the solar wind. This aerodynamic drag force accelerates slow CMEs but decelerates fast CMEs \citep[e.g.,][]{2000GeoRL..27..145G,2001SoPh..202..173V,2007SoPh..241...85V,2007A&A...472..937V,2004SoPh..221..135C,2012GeoRL..3919107S,2015ApJ...809..158S}. In this paper, the term slow (fast) refers to the condition where a CME travels slower (faster) than the solar wind which typically has a speed of $\sim$400--500~km~s$^{-1}$. Previous studies have shown that most fast CMEs reach their peak speed in the low to middle corona ($<$10~$R_\odot$) \citep[e.g.,][]{2001ApJ...559..452Z,2004ApJ...604..420Z,2007SoPh..241...85V,2010ApJ...712.1410T,2011ApJ...738..191B,2013ApJ...773..129S}. In this paper, we present an event for which the acceleration of a moderately fast CME clearly remains positive for around five hours in the high corona, and then study the mechanism acting on the continuous acceleration of the CME.

The CME speed in previous studies is usually derived by the height-time measurements in coronagraphic images where the height is located at the CME front (leading edge). However, since CMEs are known to have substantial expansion \citep[e.g.,][]{1993JGR....98.7621F,2010A&A...522A.100P,2017ApJ...848...75L,2020SoPh..295..107B}, the speed of the CME front along the propagation direction is the sum of the bulk speed indicative of the propagation of the CME center and the expansion speed relative to the center. \citet{2020SoPh..295..107B} analyzed a sample of 475 CMEs and found that the median expansion speed can reach $\sim$200--300~km~s$^{-1}$ in the corona, which is comparable to the magnitude of the bulk speed. Therefore, investigating the role of magnetic reconnection in CME dynamics by measuring the CME front variation would inevitably incorporate the effect of CME expansion. That is to say, magnetic reconnection leading to the CME true acceleration and expansion would together contribute to the speed variation at the CME front (leading edge). Furthermore, physical processes describing how magnetic reconnection contributes to CME expansion have not been investigated in depth, especially with quantitative measurements. One interpretation proposed by \citet{2004ApJ...602..422L} is that the reconnected magnetic flux accounts for the CME rapid expansion and the plasma flowing out of the current sheet fills the CME outer shell. Readers can also refer to Section \ref{sec5} for a discussion and a sketch about the broadening of the CME outer shell through the addition of plasma and magnetic flux associated with magnetic reconnection. In addition to the expansion along the radial direction, CMEs are found to undergo lateral expansion with similar magnitude to or larger than the radial expansion in the corona \citep[e.g.,][]{2005AnGeo..23.1033S,2009SoPh..260..401M,2010A&A...522A.100P,2010ApJ...724L.188P,2013ApJ...763...43C,2018ApJ...868..107V,2020SoPh..295..107B}.

In this paper, we study a CME event that occurred on 2013 February 27. The CME undergoes a long-duration acceleration in the high corona from $\sim$10 to $>$20~$R_\odot$, even though in this height range the CME speed is found to be substantially greater than the solar wind speed. Signatures of magnetic reconnection on the solar surface are found after the CME launch, which provides us a chance to study its contribution to the CME acceleration and expansion in the high corona. Section \ref{sec2} introduces the instruments and the CME event. Section \ref{sec3} studies the CME dynamics. Section \ref{sec4} analyzes the separate contribution of magnetic reconnection. Sections \ref{sec5} and \ref{sec6} provides discussion and conclusions.

\section{Observation}\label{sec2}
\subsection{Data and Instrument}\label{sec2.1}
We use the Large Angle and Spectrometric Coronagraph \citep[LASCO,][]{1995SoPh..162..357B} on board the Solar and Heliospheric Observatory \citep[SOHO,][]{domingo95} to study the CME dynamic evolution. LASCO C2 and C3 have a field of view (FOV) of 1.5--6~$R_\odot$ and 3.7--30~$R_\odot$, respectively. We also use the Sun-Earth Connection Coronal and Heliospheric Investigation \citep[SECCHI,][]{2008SSRv..136...67H} on board the ``Ahead'' and ``Behind'' satellites of the Solar TErrestial RElations Observatory \citep[STEREO,][]{2008SSRv..136....5K}, including COR1 and COR2 coronagraphs, and the Extreme Ultraviolet Imager (EUVI). During the eruption of the CME under study, STEREO-A and STEREO-B were located at 131$^\circ$ and -139$^\circ$ longitudinally in the Heliocentric Earth Ecliptic (HEE) coordinates. In situ solar wind parameters at 1~au are taken from the OMNI/Wind database.

\subsection{Event Overview}\label{sec2.2}
\begin{figure}[!hbt]
	\centering
	\includegraphics[width=0.8\textwidth]{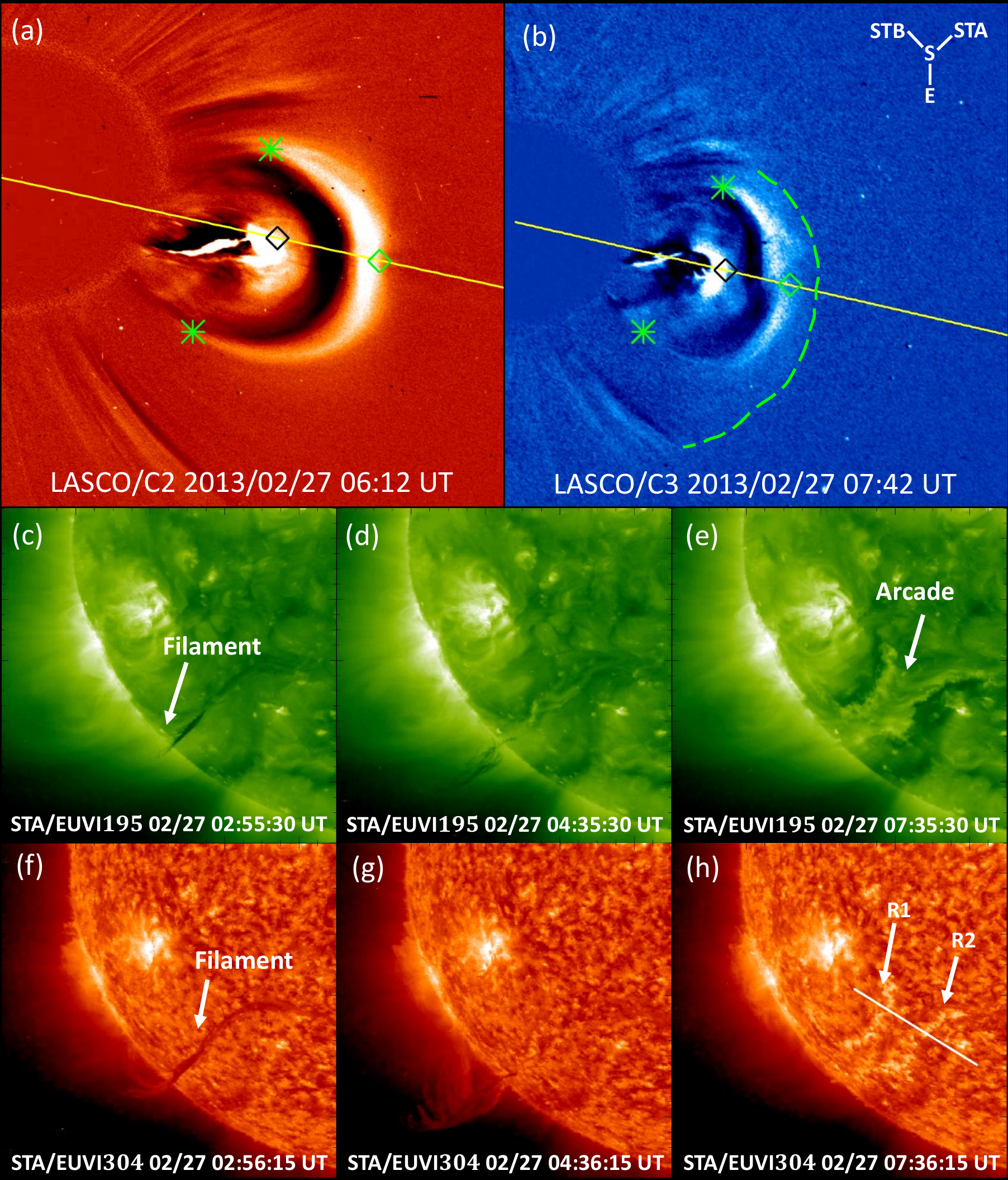}
	\caption{\small CME eruption on 2013 February 27. (a)--(b) Running-difference images of the CME in LASCO~C2 and C3 at two different times. The yellow line indicates the position angle used for the height measurement, the black and green diamonds indicate the CME core and front, and the two green asterisks locate the two lateral points. In panel (b), the dashed green curve indicates the shock front, and the inset shows the positions of STEREO-A and STEREO-B relative to the Sun (S) and Earth (E). Panels (a) and (b) show the heliocentric distance range of 1--6~$R_\odot$ and 1--17~$R_\odot$, respectively. (c)--(h) Observations of STEREO-A/EUVI 195~$\mathring{\rm{A}}$ and 304~$\mathring{\rm{A}}$. The solid line in panel (h) refers to the position of the superposed slit.} 
	\label{event}
\end{figure}

The event under study originates from the west solar limb (as seen from Earth) on 2013 February 27. Figures \ref{event}(a) and (b) show the running-difference images of the CME in LASCO C2 and C3 at two different times. The three-part structure, i.e., a bright expanding loop, followed by a relatively dark cavity and a bright core \citep{1986JGR....9110951I,2006ApJ...641..590G}, is clearly observed. The faint front ahead of the CME sharp bright front (leading edge) may refer to the shock structure \citep[e.g.,][]{2003ApJ...598.1392V}, indicated by the dashed green curve in Figure \ref{event}(b). Figures \ref{event}(c)--(h) show the observations by STEREO-A/EUVI in 195~$\mathring{\rm{A}}$ and 304~$\mathring{\rm{A}}$ wavelengths. This CME is related to a filament eruption from a quiet solar region that launches at around 04:00 UT as observed by STEREO-A/EUVI 304~$\mathring{\rm{A}}$. CME signatures in white-light appear first in LASCO C2 at 04:48 UT. Post-eruption arcades in EUVI 195~$\mathring{\rm{A}}$ and two bright ribbons in EUVI 304~$\mathring{\rm{A}}$ are observed after the CME rising. Since this CME can be observed by LASCO, STEREO-A, and STEREO-B simultaneously, we use the graduated cylindrical shell (GCS) model \citep{2006ApJ...652..763T,2009SoPh..256..111T} to obtain its 3-dimensional propagation parameters. We first focus on the propagation direction, which is S12W86 according to the GCS model. It indicates that this CME is an ideal limb event to LASCO observations, and thus we are able to directly analyze the CME dynamics in LASCO FOV with minimal projection effects.

\section{CME Dynamics}\label{sec3}
\subsection{Long-Duration CME Acceleration in the High Corona}\label{sec3.1}
\begin{figure}[!hbt]
	\centering
	\includegraphics[width=\textwidth]{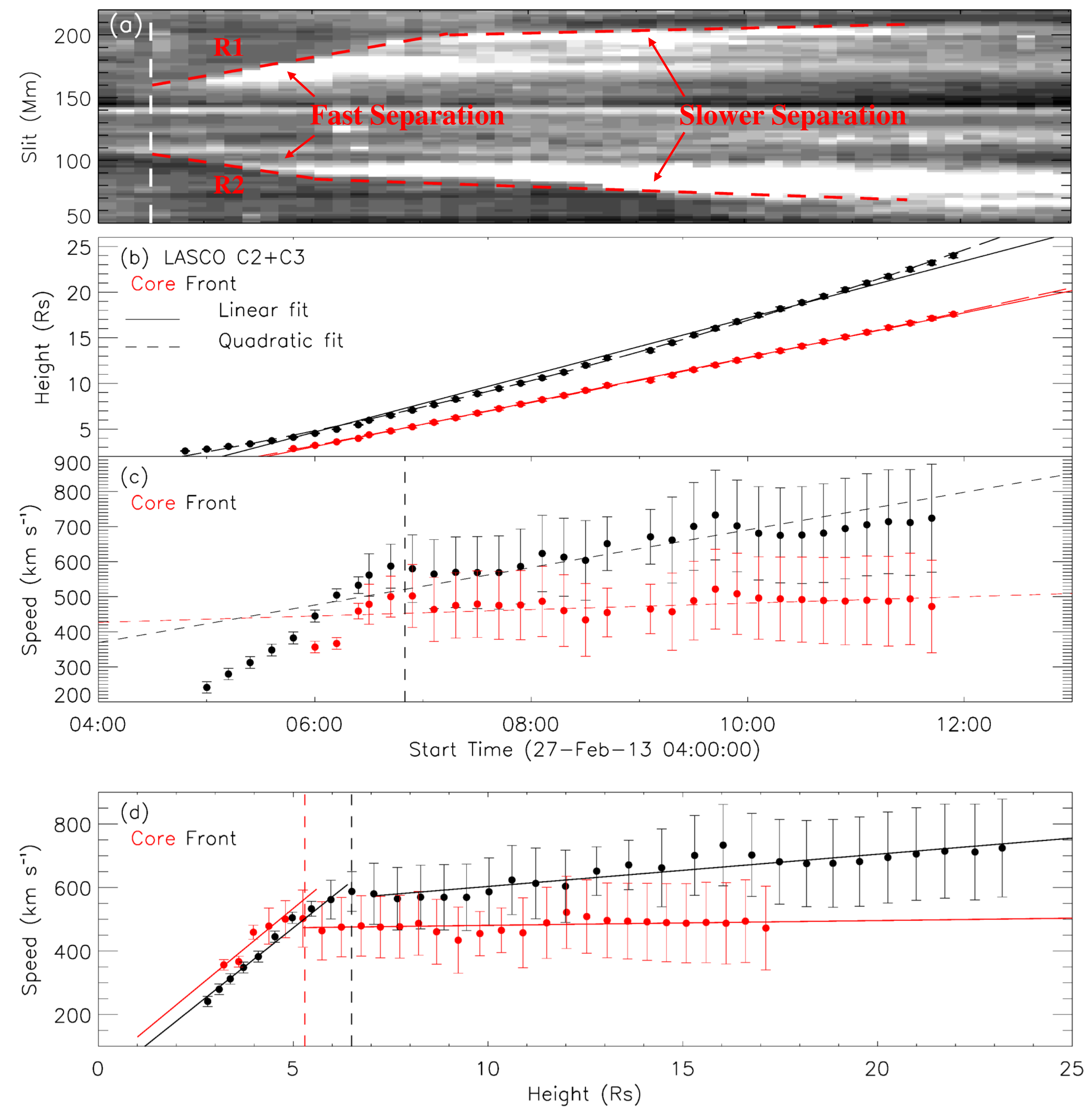}
	\caption{\small (a) Base-difference image of the slit in STEREO-A/EUVI 304~$\mathring{\rm{A}}$. Separation of the two ribbons are indicated by the dashed red lines. Vertical dashed line marks the first appearance time of the two ribbons. (b) Height-time measurement of the CME front (black) and core (red), associated with the results by a linear fit (solid line) and a quadratic fit (dashed line). (c) Speed-time measurement. The dashed lines indicate the speed obtained from the quadratic fit in panel (b). (d) Height-speed measurement. The vertical black line indicates the front height at 6.5~$R_\odot$, and the red one indicates the core height at the same time. The solid lines are the linear-fit results.} 
	\label{timevar}
\end{figure}

We carefully measure the heights of the CME front and core along the position angle of 255$^\circ$ in LASCO C2 and C3 FOV (note that S12 as derived from GCS roughly corresponds to the projected position angle of 255$^\circ$). In Figures \ref{event}(a)--(b), the yellow line indicates the selected position angle, and the green and black diamonds mark the locations of the front and core, respectively. In the following analysis, we presume that the core locates near the CME axis in LASCO FOV, and we refer to the discussion for detailed interpretations. The height-time measurements of the front and core are shown in Figure \ref{timevar}(b). Since the CME becomes increasingly fainter during its outward propagation, here we set the uncertainties in measuring the height as increasing linearly from $\pm$0.023~$R_\odot$ (i.e., plus/minus two pixels) for the initial data point in C2 image to $\pm$0.25~$R_\odot$ (i.e., plus/minus four pixels) for the final one in C3 image. 

The data points are then fitted linearly (solid line) and quadratically (dashed line). The errors below are the 1$\sigma$-uncertainty from the fits. The linear fit indicates that the CME front has an average speed significantly larger than that of the core, i.e., $602 \pm89$~km~s$^{-1}$ vs. $471 \pm20$~km~s$^{-1}$. This suggests that the CME expands radially with an average expansion speed of $131\pm 91$~km~s$^{-1}$. The quadratic fit shows that the front speed at 20~$R_\odot$ is $735 \pm26$~km~s$^{-1}$ and greater than the linearly fitted speed, and shows that the front has a positive acceleration of $a=14.0 \pm 1.0$~m~s$^{-2}$. These results are consistent with those measured by the Coordinated Data Analysis Workshop (CDAW) CME catalog, but the magnitudes are slightly smaller than those in the catalog because the catalog measures the relatively faint front of the shock structure. As for the core, the quadratic-fit speed is similar to the linear-fit speed, i.e., $507 \pm20$~km~s$^{-1}$ vs. $471 \pm20$~km~s$^{-1}$, associated with a smaller acceleration of $a=2.5 \pm 0.6$~m~s$^{-2}$.

We then obtain the speed-time and speed-height variations in Figures \ref{timevar}(c) and (d), respectively. The speed at each time step is derived by a linear fit to three successive height-time measurements, and thus there are no speed data points for the first and final time steps. The dashed lines in Figure \ref{timevar}(c) show the speed derived by the quadratic fit to the height-time measurements, which indicate a linearly increased front speed and a constant core speed. It is also found that the front and the core experience two propagation phases, i.e., an impulsive acceleration (Phase-I) followed by a constant-speed propagation for the core and a smooth acceleration for the front (Phase-II). Based on the change of the CME front speed, the two phases are separated by the time at $\sim$06:50 UT and the height of $\sim$6.5~$R_\odot$, which are indicated by the vertical dashed black line in Figures \ref{timevar}(c) and \ref{timevar}(d). Vertical dashed red line in Figure \ref{timevar}(d) indicates the height of the core at 06:50 UT. The speed-time and speed-height data points in the two phases are then linearly fitted, and shown by the solid lines in Figure \ref{timevar}(d). In Phase I, $a=59.1 \pm 3.7$~m~s$^{-2}$ for the front, and $a=65.2 \pm11.4$~m~s$^{-2}$ for the core. The acceleration magnitudes are the same for both the core and the front. In Phase II, $a=10.1 \pm4.7$~m~s$^{-2}$ for the front, much higher than $a=1.0 \pm 0.7$~m~s$^{-2}$ for the core, indicating that in this height range, the CME front continues accelerating while its core can be approximated as propagating with a constant speed.

We focus on Phase-II, and emphasize here again that this long-duration acceleration is seen albeit the speed of the front at the start of Phase II is $\sim$600~km~s$^{-1}$ and larger than the expected solar wind speed at this height of 6.5~$R_\odot$ (see Section \ref{sec4} for the solar wind speed estimation). The acceleration of the CME front here lasts for nearly five hours as the CME propagates in the high corona. To the best of our knowledge, this is the first reported instance of a CME that continues to accelerate in the high corona, and even to $>$20~$R_\odot$, though it is already faster than the solar wind. Although the acceleration is not as large as compared to that in Phase-I, the long acceleration duration leads to: (1) the front speed increases by $\sim$170~km~s$^{-1}$ as the CME propagates in the high corona, and (2) the ``final'' speed reaches 700~km~s$^{-1}$ and above. Furthermore, the comparison between the acceleration of the front and core indicates that the apparent acceleration in observations of the CME front in Phase-II is primarily caused by a radial expansion with an acceleration of $a_{meas} = 9.1 \pm 4.8$~m~s$^{-2}$. If not for this expansion, the core would accelerate at the same rate due to the frozen-in conditions in the CME plasma.

Looking at the uncertainties in the CME speed, the finding about the continuous acceleration of the CME in the high corona may be questioned. To confirm that the main result holds, we perform the following calculations. Since the uncertainties come from the height measurement in coronagraphs, we obtain the acceleration by quadratically fitting the height-time measurements for the CME starting from larger distances, e.g., 15 and 20~$R_\odot$. These fitted acceleration is still around 10~m~s$^{-2}$. Furthermore, since the CME front becomes fainter as it propagates outward while the true leading edge might not be discernible, the height of the located front in coronagraphs may be lower than the true one. These two points would further support that the continuous acceleration of the CME front is true in the high corona. In addition, the CME front is too faint to be distinguished above 25~$R_\odot$ which may bring even larger uncertainties, and thus it is still unclear whether this CME has reached its final speed at this height.

\subsection{CME Lateral Expansion}
\begin{figure}[!hbt]
	\centering
	\includegraphics[width=\textwidth]{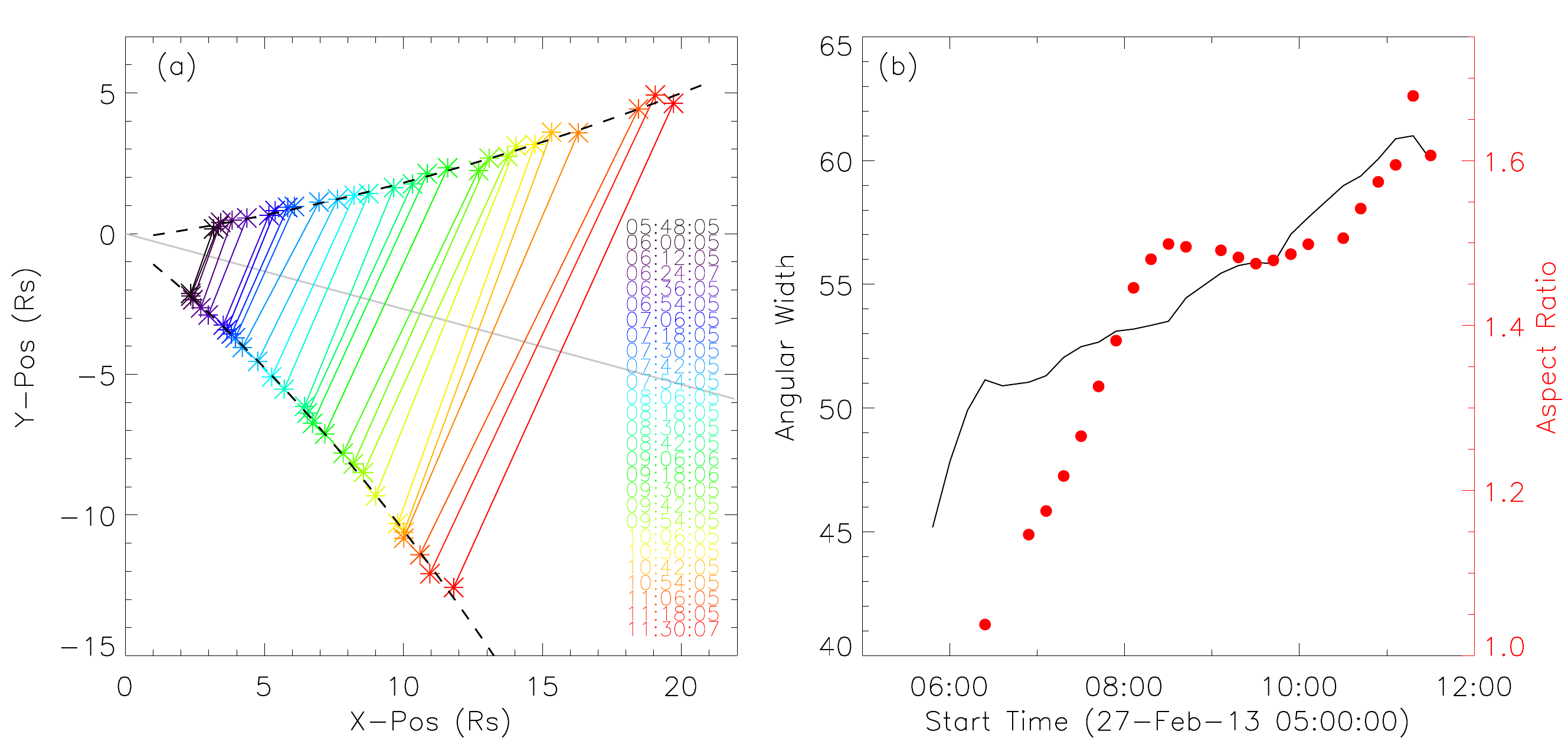}
	\caption{\small (a) Variation of the CME lateral width at different times in X-Y plane. The black solid line indicates the position angle of 255$^\circ$, and the two curves show the quadratic fit to the X-Y data points. (b) Variation of the CME angular width (black) and the aspect ratio (red).}  
	\label{width}
\end{figure}

We turn our attention to the magnitude of the CME lateral expansion which should occur jointly with the radial expansion. The CME lateral width is calculated as the length of the line connecting the two asterisks as shown in Figures \ref{event}(a) and (b). The two asterisks at each time step mark the outermost locations of both CME side-edges. Figure \ref{width}(a) shows the variations of the lateral length associated with the asterisks at different times (varying with colors) in X-Y plane, where the origin indicates the center of the Sun. The solid black line indicates the position angle of $255^\circ$, which is found to be approximately perpendicular to those lateral lines. The two dashed curves correspond to the quadratic fit to the upper and lower sets of asterisks. A linear fit to half of the lateral length shows that the CME lateral expansion at one edge side has an average speed of $260 \ (\pm 17)$~km~s$^{-1}$, while a quadratic fit reveals that the acceleration of the expansion at one side is $14.7 \ (\pm 2.3)$~m~s$^{-2}$. These indicate that the CME has a stronger lateral expansion compared to the radial expansion, and such a positive acceleration confirms that the lateral expansion occurs jointly with the radial one.

Figure \ref{width}(b) shows the variation of the angular width ($\omega$) which is the total width between the two asterisks. It is found that $\omega$ increases by $\sim$10$^\circ$ in Phase-II, to reach $\sim$60$^\circ$ at $\sim$20~$R_\odot$. Again, the CME might continue to expand laterally beyond 20~$R_\odot$. The aspect ratio, i.e., the ratio of half of the lateral length to the radial length from the core to the front, is shown in Figure \ref{width}(b). The aspect ratio increases impulsively at first, and then keeps approximately constant at 1.5--1.6. A ratio greater than unity indicates that the cross-section of the CME is an ellipse with the major axis in the lateral direction and the minor axis in the radial direction. The aspect ratio $>1$ is consistent with the above results that the speed and acceleration in the lateral direction are larger than those in the radial direction.

\subsection{Magnetic Reconnection Associated with the CME Eruption}\label{sec-rec}
Signatures of magnetic reconnection on the solar surface are visible after the CME eruption. Two ribbons start brightening up at around 05:00 UT in STEREO-A/EUVI 304~$\mathring{\rm{A}}$ images, associated with a post-eruption arcade in 195~$\mathring{\rm{A}}$ images, as shown in Figures \ref{event}(e) and (h), respectively. The two ribbons separate as the eruption proceeds, shown by the stack plot of a virtual slit (Figure \ref{event}(h)) across the ribbons (with solar rotation corrected) in Figure \ref{timevar}(a). Ribbon separation acts as an indicator for the ongoing magnetic reconnection processes \citep[e.g.,][]{2002A&ARv..10..313P}. In Figure \ref{timevar}(a), vertical dashed line marks the first appearance time of the two ribbons. Dashed red lines locate the ribbon fronts which are visually identified, and the separation speed can be deduced by the associated slope. The separation of the ribbons is fast at first before 06:00--07:00 UT, which is consistent with the CME impulsive acceleration in Phase-I. The separation then slows down but lasts for around five hours to around 12:00 UT for which the signature is clearer as seen by the front of the ribbon R2, and its duration is comparable to the CME acceleration period in Phase-II. Since magnetic reconnection plays a significant role in the CME eruption processes, the separation motion of the two ribbons indicates that for the current event reconnection processes contribute to the CME dynamic evolution, impulsively from the low to middle corona, and continuously for several hours in the high corona. We come back to the consequence of this reconnection in the discussion.

\section{Contribution of Magnetic Reconnection to CME Dynamics}\label{sec4}
\begin{figure}[!hbt]
	\centering
	\includegraphics[width=0.9\textwidth]{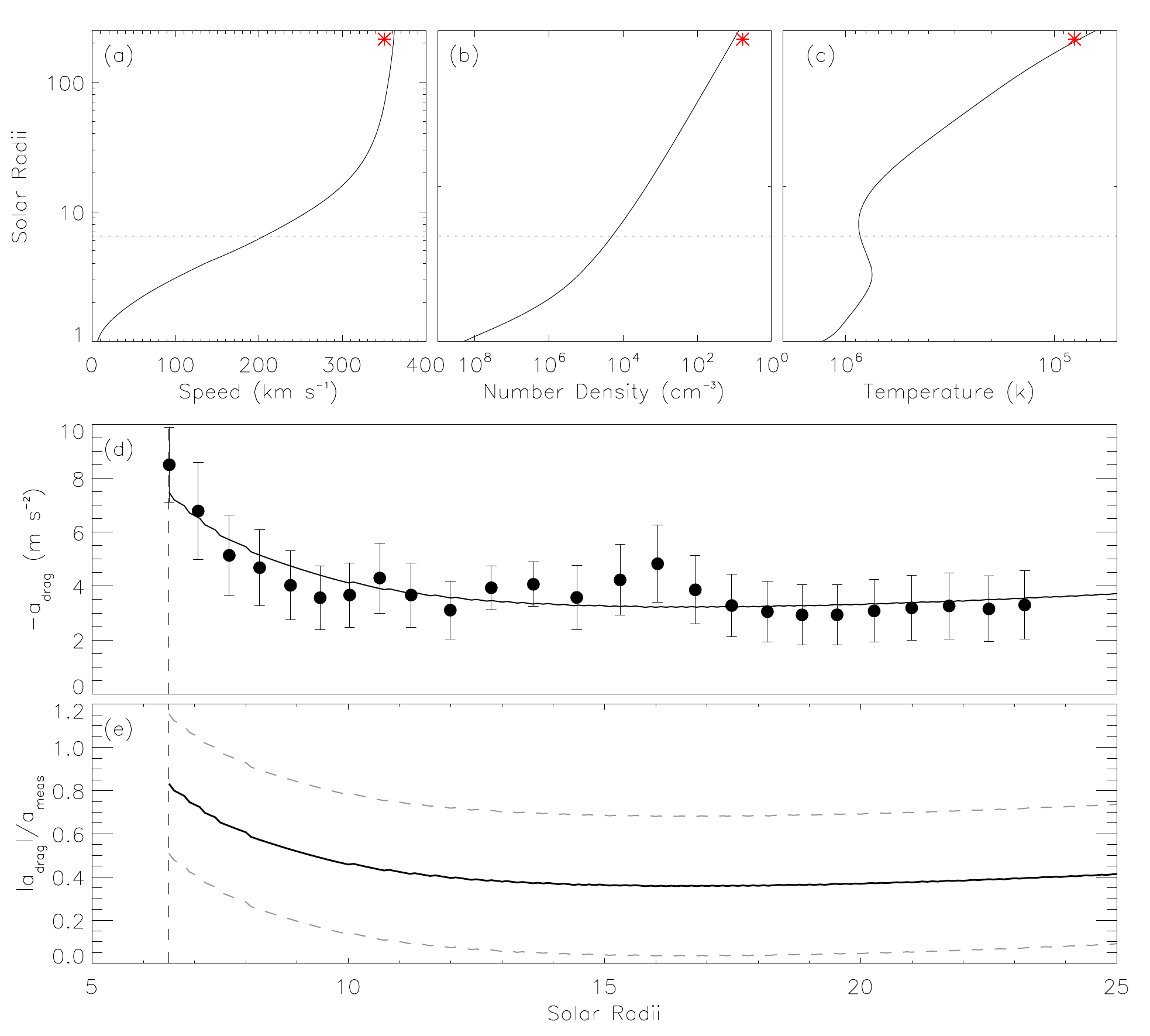}
	\caption{\small (a)--(c) Variation of the simulated solar wind proton speed, number density and temperature with heliocentric distance, compared with the in-situ observational data at 1~au (red asterisks). Horizontal dotted line indicates the distance of 6.5~$R_\odot$. (d) $a_{drag}$ in the drag-based model by using the CME front speed based on discrete height-time measurements (filled circle) and quadratical fit to the height-time measurements (solid curve). (e) Ratio of $|a_{drag}|$ to $a_{meas}$. The two dashed curves indicate the lower and upper uncertainties. The vertical dashed lines in panels (d) and (e) indicate the height at 6.5~$R_\odot$.}  
	\label{drag}
\end{figure}

As presented in Section \ref{sec3.1}, the apparent acceleration in observations of the CME front is related to the CME expansion. However, as fast CMEs may experience deceleration due to the solar wind drag, a true CME acceleration is required to compensate for the effect of the drag and be present for the CME not to decelerate. Consequently, magnetic reconnection simultaneously contributes to both the true acceleration and the apparent acceleration (related to the expansion). The true acceleration for our event equals to the magnitude of the deceleration caused by the drag. We then use a drag-based model \citep[see e.g.,][]{1996JGR...101.4855C,2004SoPh..221..135C,2007A&A...472..937V,2012GeoRL..3919107S} to estimate the magnitude of the drag effect, which is expressed by 
\begin{equation}\label{dragm}
	a_{drag}=\frac{F_{drag}}{m_{CME}}=-\frac{C_D A_{CME} n_p m_p }{2 m_{CME}}(V_{CME}-V_{SW})|V_{CME}-V_{SW}|
\end{equation}
and depends on the CME and solar wind characteristics. $C_D$ is the dimensionless drag coefficient and here we use a hybrid viscosity model for $C_D$ as presented by \citet{2012GeoRL..3919107S}. $A_{CME}$ is the CME cross-sectional area, and $A_{CME}\sim \pi R_{CME}^2$ where $R_{CME}$ approximately equals to half of the CME lateral length. $m_p$ and $m_{CME}$ are the proton and CME mass, respectively. In this paper, $m_{CME}$ is measured when the CME front is at the height of $\sim$10~$R_\odot$ in the C3 calibrated image by the solar software (SSW) program of CME$\_$MASS.pro, and the value is around $1.2\times 10^{15}$~g. One can also refer to \citet{2000ApJ...534..456V} for details about the mass measurement procedure. $V_{CME}$ is the CME speed, obtained by using the CME front speed as shown in Figure \ref{timevar}(d). $n_p$ is the solar wind proton number density, and $V_{SW}$ is the solar wind speed. In this paper, we use an 1-dimensional magnetohydrodynamic (MHD) numerical simulation \citep{1997JGR...10214661H} to estimate the solar wind parameters which are dependent on the heliocentric distance. The simulations are adjusted to ensure the simulated results at 1~au distance are as close as possible to the in situ observations obtained from OMNI/Wind database. Figures \ref{drag}(a)--(c) show the evolution of the simulated speed, proton number density, and temperature of the solar wind, while the in situ measurements are marked by the red asterisks. We note that panel (a) further confirms that the CME is faster than the solar wind in the corona, e.g., the solar wind speed is around 200--300~km~s$^{-1}$ in the height range of 6.5--25~$R_\odot$.

The estimated $a_{drag}$ is shown in Figure \ref{drag}(d), in which the filled circles are the results obtained by using the CME front speed from the discrete height-time measurements, and the solid curve is obtained by using the front speed from the quadratic fit of the height-time measurements in Phase-II. The uncertainty of $a_{drag}$ in the figure comes from the uncertainty in measuring the CME speed. Uncertainty in the CME mass measurement is discussed later. $|a_{drag}|$ is found to decrease from $\sim$9 to $\sim$4 $\rm{m \ s^{-2}}$ during the CME propagation in the high corona. Figure \ref{drag}(e) shows the ratio of $|a_{drag}|$ (using the values of the solid curve in (d)) to $a_{meas}$ (9.1 $\rm{m \ s^{-2}}$) by the solid curve, with uncertainties shown by the two dashed curves. It is found that this ratio decreases from about 0.8 to 0.4, suggesting that in Phase-II the contribution of magnetic reconnection to the apparent acceleration related to the expansion is comparable to or even stronger than its contribution to the true acceleration that compensates the drag for the CME.

\section{Discussion}\label{sec5}
We first return to our assumption that the CME bright core corresponds to the CME center. It is traditionally suggested that the CME bright core in coronagraphic images corresponds to the prominence material. The dense prominence/filament material is suspended in the magnetic dips at the bottom of a flux rope \citep{2006ApJ...641..590G}, leading to a separation between the prominence/filament and the CME axis/center. However, this scenario might be primarily true during the CME initiation phase; when the CME propagates into the mid and outer corona, the bright core can be approximately co-located with the CME center. There are three arguments supporting this assumption. First, \citet{2017ApJ...834...86H} and \citet{2018ApJ...868..107V} presented observations showing that the inner core of the three-part coronagraph CME is not necessarily a prominence, while \citet{2018ApJ...868..107V} argued that the core is still a flux rope structure. Second, the large amount of newly added flux to the core (or initial flux rope) through magnetic reconnection during the eruption \citep[e.g.,][]{2000JGR...105.2375L,2004ApJ...602..422L} would keep the core in the center of the erupting structure. Third, the variation of the aspect ratio shown in Figure \ref{width} is relatively consistent with expectation \citep[e.g.,][]{2011ApJ...731..109S} that the CME starts with a circular cross-section. If the core is located at the back of the CME, the initial shape would already be very elliptical, which is contrary to past studies \citep[e.g.,][]{2001ApJ...562.1045K,2010A&A...522A.100P,2013ApJ...763...43C}. Note that if the core is not the center of the flux rope, this will affect the quantitative result but not the acceleration issue itself. The CME front is still seen as accelerating to large speeds but the acceleration related to expansion would only be half of what we consider here. The acceleration of the CME front would still need to be explained.

\begin{figure}[!hbt]
	\centering
	\includegraphics[width=\textwidth]{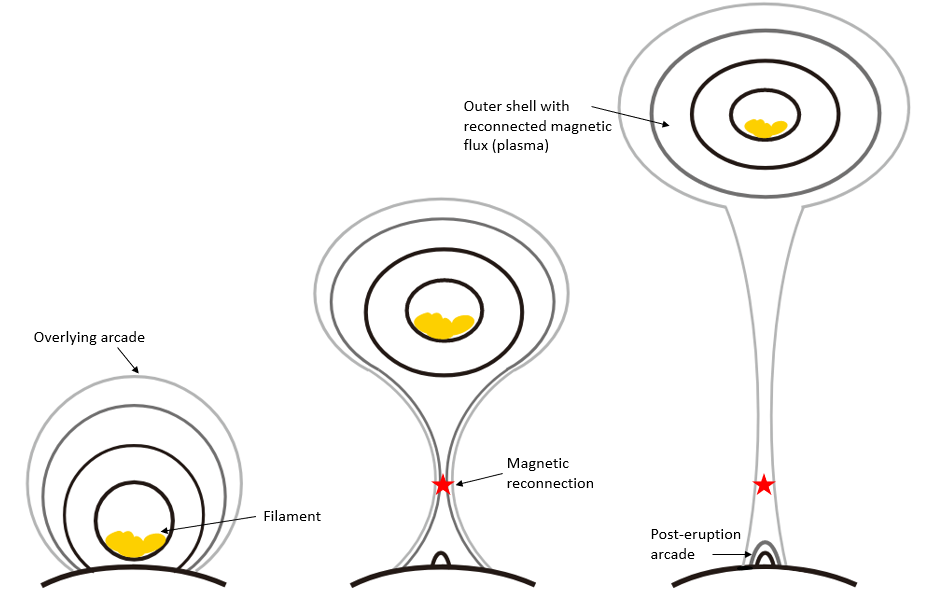}
	\caption{\small Sketch of a CME eruption associated with magnetic reconnection processes. Filament material is shown by the orange shadow, and magnetic reconnection site is marked by the red asteroid. The shape of the cross section of the flux rope is designed to vary from circular to elliptical to be consistent with the variation of the aspect ratio as found in Figure \ref{width}(b).}  
	\label{sketch}
\end{figure}

Figure \ref{sketch} shows a sketch of the CME eruption in this paper associated with a filament in the core region of the CME flux rope. This figure only shows the cross-section structures, and the shape of the cross section of the flux rope is designed to vary from circular to elliptical to be consistent with the aspect ratio variation as shown in Figure \ref{width}(b). In the left panel of the pre-eruption phase, the filament with heavy material (orange shadow) is suspended at the bottom of the initial flux rope. The flux rope then erupts associated with ongoing magnetic reconnection processes, as shown in the middle and right panels. Magnetic reconnection adds additional (poloidal) magnetic fluxes to the initial flux rope and forms the outer shells, which contributes to an additional expansion. Note that we do not show the normal expansion as the CME propagates outward, but emphasize the role of reconnection in the expansion, similar to Figure 11 in \citet{2004ApJ...602..422L}. Post-eruption arcade associated with reconnection is formed beneath the flux rope, as also seen in Figure \ref{event}(e) for the CME event. As the CME propagates upward, gravity decreases significantly, and the initial flux rope has been enveloped by a large amount of reconnected (poloidal) magnetic fluxes. The decreased gravity and the strong magnetic tension force from reconnected poloidal fluxes may keep the initial flux rope associated with the filament (which corresponds to the CME bright core in coronagraph images) near the center of the CME. In addition, this scenario does not consider the mass-unloading of the filament material, which was found to facilitate the CME radial expansion \citep{2018SoPh..293....7J}.

We next turn our attention to the uncertainty in the CME mass measurement. Accurate measurements are important, because (1) the calculation of $a_{drag}$ is sensitive to CME mass as clearly seen in Equation \ref{dragm}, and (2) the pile-up of CME mass \citep[e.g.,][]{2005ApJ...627.1019L,2015ApJ...812...70F} may bring additional uncertainties as we set the mass as constant in the drag-based model. However, \citet{2013ApJ...768...31B} found that the mass pile-up is rare by analyzing the evolution of the deprojected CME mass, and \citet{2018SoPh..293...55H} also argued that there is no observed pile-up for CMEs in LASCO FOV by studying the variation of the electron density of the CME front. In addition, we test the influence of an $\pm15\%$ uncertainty of the measured mass on the comparison between CME expansion and acceleration. Still, we find similar contributions of reconnection to the expansion and acceleration.

\section{Conclusion}\label{sec6}
The CME on 2013 February 27, originating from a solar quiet region, is found to be associated with a continuous acceleration of its front for around five hours in the high corona and even up to 20~$R_\odot$, though its speed has already reached $\sim$700~km~s$^{-1}$, i.e., larger than the solar wind speed. This indicates that even fast CME events may not reach their ``final'' (or maximum) speed at 20~$R_\odot$. The long-duration acceleration has important consequences for some heliospheric models \citep[e.g.,][]{2003AdSpR..32..497O,2018JSWSC...8A..35P,2019A&A...626A.122S}, where the inner boundary is set at $\sim$20~$R_\odot$. We find that the long-duration apparent acceleration of the CME front is caused by the CME's own expansion, which further suggests the importance of considering expansion in studying CME dynamics and in those heliospheric models that usually assume a uniform speed distribution within the CME. For a discussion of including expansion in the initiation of CME at 20~$R_\odot$ in a heliospheric model, see \citet{2019A&A...626A.122S}. A long-duration magnetic reconnection occurs after the eruption, which contributes to the evolution of the CME dynamics. For this CME especially when it is in the high corona, magnetic reconnection results in (1) an apparent acceleration related to the expansion and (2) a true acceleration compensating the deceleration caused by the solar wind drag. Based on quantitative comparison between the CME expansion and true acceleration in the high corona, it indicates that the contribution of reconnection to the CME expansion is comparable to the contribution to the CME true acceleration in that height range. In addition, a study about the Alfv\'{e}n point locating up to 12--15~$R_\odot$ \citep{2014ApJ...787..124D}, and a recent paper showing that the Alfv\'{e}n point (or surface) even reaching as high as $\sim$19~$R_\odot$ using the observations from the Parker Solar Probe \citep{2021PhRvL.127y5101K}, support the possibility of magnetic reconnection continuously contributing to the CME dynamics from the low to high corona.


\textbf{Acknowledgement} We acknowledge the use of the data from SOHO, STEREO and OMNI/Wind. SOHO is a mission of international cooperation between ESA and NASA. Research for this work was made possible by NASA grants 80NSSC19K0831, 80NSSC20K0700, and 80NSSC20K0431. 

\clearpage

\bibliographystyle{apj}
\bibliography{cmerec}

\begin{thebibliography}{}
\expandafter\ifx\csname natexlab\endcsname\relax\def\natexlab#1{#1}\fi

\bibitem[{{Balmaceda} {et~al.}(2020){Balmaceda}, {Vourlidas}, {Stenborg}, \&
  {St. Cyr}}]{2020SoPh..295..107B}
{Balmaceda}, L.~A., {Vourlidas}, A., {Stenborg}, G., \& {St. Cyr}, O.~C. 2020,
  \solphys, 295, 107

\bibitem[{{Bein} {et~al.}(2011){Bein}, {Berkebile-Stoiser}, {Veronig},
  {Temmer}, {Muhr}, {Kienreich}, {Utz}, \&
  {Vr{\v{s}}nak}}]{2011ApJ...738..191B}
{Bein}, B.~M., {Berkebile-Stoiser}, S., {Veronig}, A.~M., {et~al.} 2011, \apj,
  738, 191

\bibitem[{{Bein} {et~al.}(2013){Bein}, {Temmer}, {Vourlidas}, {Veronig}, \&
  {Utz}}]{2013ApJ...768...31B}
{Bein}, B.~M., {Temmer}, M., {Vourlidas}, A., {Veronig}, A.~M., \& {Utz}, D.
  2013, \apj, 768, 31

\bibitem[{{Brueckner} {et~al.}(1995){Brueckner}, {Howard}, {Koomen},
  {Korendyke}, {Michels}, {Moses}, {Socker}, {Dere}, {Lamy}, {Llebaria},
  {Bout}, {Schwenn}, {Simnett}, {Bedford}, \& {Eyles}}]{1995SoPh..162..357B}
{Brueckner}, G.~E., {Howard}, R.~A., {Koomen}, M.~J., {et~al.} 1995, \solphys,
  162, 357

\bibitem[{{Cargill}(2004)}]{2004SoPh..221..135C}
{Cargill}, P.~J. 2004, \solphys, 221, 135

\bibitem[{{Cargill} {et~al.}(1996){Cargill}, {Chen}, {Spicer}, \&
  {Zalesak}}]{1996JGR...101.4855C}
{Cargill}, P.~J., {Chen}, J., {Spicer}, D.~S., \& {Zalesak}, S.~T. 1996, \jgr,
  101, 4855

\bibitem[{{Cheng} {et~al.}(2013){Cheng}, {Zhang}, {Ding}, {Liu}, \&
  {Poomvises}}]{2013ApJ...763...43C}
{Cheng}, X., {Zhang}, J., {Ding}, M.~D., {Liu}, Y., \& {Poomvises}, W. 2013,
  \apj, 763, 43

\bibitem[{{DeForest} {et~al.}(2014){DeForest}, {Howard}, \&
  {McComas}}]{2014ApJ...787..124D}
{DeForest}, C.~E., {Howard}, T.~A., \& {McComas}, D.~J. 2014, \apj, 787, 124

\bibitem[{{Domingo} {et~al.}(1995){Domingo}, {Fleck}, \& {Poland}}]{domingo95}
{Domingo}, V., {Fleck}, B., \& {Poland}, A.~I. 1995, 162, 1

\bibitem[{{Farrugia} {et~al.}(1993){Farrugia}, {Burlaga}, {Osherovich},
  {Richardson}, {Freeman}, {Lepping}, \& {Lazarus}}]{1993JGR....98.7621F}
{Farrugia}, C.~J., {Burlaga}, L.~F., {Osherovich}, V.~A., {et~al.} 1993, \jgr,
  98, 7621

\bibitem[{{Feng} {et~al.}(2015){Feng}, {Wang}, {Shen}, {Shen}, {Inhester},
  {Lu}, \& {Gan}}]{2015ApJ...812...70F}
{Feng}, L., {Wang}, Y., {Shen}, F., {et~al.} 2015, \apj, 812, 70

\bibitem[{{Forbes}(2000)}]{2000JGR...10523153F}
{Forbes}, T.~G. 2000, \jgr, 105, 23153

\bibitem[{{Gibson} {et~al.}(2006){Gibson}, {Foster}, {Burkepile}, {de Toma}, \&
  {Stanger}}]{2006ApJ...641..590G}
{Gibson}, S.~E., {Foster}, D., {Burkepile}, J., {de Toma}, G., \& {Stanger}, A.
  2006, \apj, 641, 590

\bibitem[{{Gopalswamy} {et~al.}(2000){Gopalswamy}, {Lara}, {Lepping}, {Kaiser},
  {Berdichevsky}, \& {St. Cyr}}]{2000GeoRL..27..145G}
{Gopalswamy}, N., {Lara}, A., {Lepping}, R.~P., {et~al.} 2000, \grl, 27, 145

\bibitem[{{Gou} {et~al.}(2020){Gou}, {Veronig}, {Liu}, {Zhuang},
  {Dumbovi{\'c}}, {Podladchikova}, {Reid}, {Temmer}, {Dissauer},
  {Vr{\v{s}}nak}, \& {Wang}}]{2020ApJ...897L..36G}
{Gou}, T., {Veronig}, A.~M., {Liu}, R., {et~al.} 2020, \apjl, 897, L36

\bibitem[{{Howard} \& {Vourlidas}(2018)}]{2018SoPh..293...55H}
{Howard}, R.~A., \& {Vourlidas}, A. 2018, \solphys, 293, 55

\bibitem[{{Howard} {et~al.}(2008){Howard}, {Moses}, {Vourlidas}, {Newmark},
  {Socker}, {Plunkett}, {Korendyke}, {Cook}, {Hurley}, {Davila}, {Thompson},
  {St Cyr}, {Mentzell}, {Mehalick}, {Lemen}, {Wuelser}, {Duncan}, {Tarbell},
  {Wolfson}, {Moore}, {Harrison}, {Waltham}, {Lang}, {Davis}, {Eyles},
  {Mapson-Menard}, {Simnett}, {Halain}, {Defise}, {Mazy}, {Rochus}, {Mercier},
  {Ravet}, {Delmotte}, {Auchere}, {Delaboudiniere}, {Bothmer}, {Deutsch},
  {Wang}, {Rich}, {Cooper}, {Stephens}, {Maahs}, {Baugh}, {McMullin}, \&
  {Carter}}]{2008SSRv..136...67H}
{Howard}, R.~A., {Moses}, J.~D., {Vourlidas}, A., {et~al.} 2008, \ssr, 136, 67

\bibitem[{{Howard} {et~al.}(2017){Howard}, {DeForest}, {Schneck}, \&
  {Alden}}]{2017ApJ...834...86H}
{Howard}, T.~A., {DeForest}, C.~E., {Schneck}, U.~G., \& {Alden}, C.~R. 2017,
  \apj, 834, 86

\bibitem[{{Hu} {et~al.}(1997){Hu}, {Esser}, \& {Habbal}}]{1997JGR...10214661H}
{Hu}, Y.~Q., {Esser}, R., \& {Habbal}, S.~R. 1997, \jgr, 102, 14661

\bibitem[{{Illing} \& {Hundhausen}(1986)}]{1986JGR....9110951I}
{Illing}, R.~M.~E., \& {Hundhausen}, A.~J. 1986, \jgr, 91, 10951

\bibitem[{{Jenkins} {et~al.}(2018){Jenkins}, {Long}, {van Driel-Gesztelyi}, \&
  {Carlyle}}]{2018SoPh..293....7J}
{Jenkins}, J.~M., {Long}, D.~M., {van Driel-Gesztelyi}, L., \& {Carlyle}, J.
  2018, \solphys, 293, 7

\bibitem[{{Kaiser} {et~al.}(2008){Kaiser}, {Kucera}, {Davila}, {St. Cyr},
  {Guhathakurta}, \& {Christian}}]{2008SSRv..136....5K}
{Kaiser}, M.~L., {Kucera}, T.~A., {Davila}, J.~M., {et~al.} 2008, \ssr, 136, 5

\bibitem[{{Kasper} {et~al.}(2021){Kasper}, {Klein}, {Lichko}, {Huang}, {Chen},
  {Badman}, {Bonnell}, {Whittlesey}, {Livi}, {Larson}, {Pulupa}, {Rahmati},
  {Stansby}, {Korreck}, {Stevens}, {Case}, {Bale}, {Maksimovic}, {Moncuquet},
  {Goetz}, {Halekas}, {Malaspina}, {Raouafi}, {Szabo}, {MacDowall}, {Velli},
  {Dudok de Wit}, \& {Zank}}]{2021PhRvL.127y5101K}
{Kasper}, J.~C., {Klein}, K.~G., {Lichko}, E., {et~al.} 2021, \prl, 127, 255101

\bibitem[{Kilpua {et~al.}(2019)Kilpua, Lugaz, Mays, \& Temmer}]{kilpua2019}
Kilpua, E., Lugaz, N., Mays, M.~L., \& Temmer, M. 2019, Space Weather, 17, 498

\bibitem[{{Krall} {et~al.}(2001){Krall}, {Chen}, {Duffin}, {Howard}, \&
  {Thompson}}]{2001ApJ...562.1045K}
{Krall}, J., {Chen}, J., {Duffin}, R.~T., {Howard}, R.~A., \& {Thompson}, B.~J.
  2001, \apj, 562, 1045

\bibitem[{{Lamy} {et~al.}(2019){Lamy}, {Floyd}, {Boclet}, {Wojak}, {Gilardy},
  \& {Barlyaeva}}]{2019SSRv..215...39L}
{Lamy}, P.~L., {Floyd}, O., {Boclet}, B., {et~al.} 2019, \ssr, 215, 39

\bibitem[{{Lin} \& {Forbes}(2000)}]{2000JGR...105.2375L}
{Lin}, J., \& {Forbes}, T.~G. 2000, \jgr, 105, 2375

\bibitem[{{Lin} {et~al.}(2004){Lin}, {Raymond}, \& {van
  Ballegooijen}}]{2004ApJ...602..422L}
{Lin}, J., {Raymond}, J.~C., \& {van Ballegooijen}, A.~A. 2004, \apj, 602, 422

\bibitem[{{Lugaz} {et~al.}(2017){Lugaz}, {Farrugia}, {Winslow}, {Small},
  {Manion}, \& {Savani}}]{2017ApJ...848...75L}
{Lugaz}, N., {Farrugia}, C.~J., {Winslow}, R.~M., {et~al.} 2017, \apj, 848, 75

\bibitem[{{Lugaz} {et~al.}(2005){Lugaz}, {Manchester}, \&
  {Gombosi}}]{2005ApJ...627.1019L}
{Lugaz}, N., {Manchester}, W.~B., I., \& {Gombosi}, T.~I. 2005, \apj, 627, 1019

\bibitem[{{Michalek} {et~al.}(2009){Michalek}, {Gopalswamy}, \&
  {Yashiro}}]{2009SoPh..260..401M}
{Michalek}, G., {Gopalswamy}, N., \& {Yashiro}, S. 2009, \solphys, 260, 401

\bibitem[{{Michalek} {et~al.}(2015){Michalek}, {Gopalswamy}, {Yashiro}, \&
  {Bronarska}}]{2015SoPh..290..903M}
{Michalek}, G., {Gopalswamy}, N., {Yashiro}, S., \& {Bronarska}, K. 2015,
  \solphys, 290, 903

\bibitem[{{Odstrcil}(2003)}]{2003AdSpR..32..497O}
{Odstrcil}, D. 2003, Advances in Space Research, 32, 497

\bibitem[{{Patsourakos} {et~al.}(2010{\natexlab{a}}){Patsourakos}, {Vourlidas},
  \& {Kliem}}]{2010A&A...522A.100P}
{Patsourakos}, S., {Vourlidas}, A., \& {Kliem}, B. 2010{\natexlab{a}}, \aap,
  522, A100

\bibitem[{{Patsourakos} {et~al.}(2010{\natexlab{b}}){Patsourakos}, {Vourlidas},
  \& {Stenborg}}]{2010ApJ...724L.188P}
{Patsourakos}, S., {Vourlidas}, A., \& {Stenborg}, G. 2010{\natexlab{b}},
  \apjl, 724, L188

\bibitem[{{Pomoell} \& {Poedts}(2018)}]{2018JSWSC...8A..35P}
{Pomoell}, J., \& {Poedts}, S. 2018, Journal of Space Weather and Space
  Climate, 8, A35

\bibitem[{{Priest} \& {Forbes}(2002)}]{2002A&ARv..10..313P}
{Priest}, E.~R., \& {Forbes}, T.~G. 2002, \aapr, 10, 313

\bibitem[{{Sachdeva} {et~al.}(2015){Sachdeva}, {Subramanian}, {Colaninno}, \&
  {Vourlidas}}]{2015ApJ...809..158S}
{Sachdeva}, N., {Subramanian}, P., {Colaninno}, R., \& {Vourlidas}, A. 2015,
  \apj, 809, 158

\bibitem[{{Savani} {et~al.}(2011){Savani}, {Owens}, {Rouillard}, {Forsyth},
  {Kusano}, {Shiota}, \& {Kataoka}}]{2011ApJ...731..109S}
{Savani}, N.~P., {Owens}, M.~J., {Rouillard}, A.~P., {et~al.} 2011, \apj, 731,
  109

\bibitem[{{Schwenn} {et~al.}(2005){Schwenn}, {dal Lago}, {Huttunen}, \&
  {Gonzalez}}]{2005AnGeo..23.1033S}
{Schwenn}, R., {dal Lago}, A., {Huttunen}, E., \& {Gonzalez}, W.~D. 2005,
  Annales Geophysicae, 23, 1033

\bibitem[{{Scolini} {et~al.}(2019){Scolini}, {Rodriguez}, {Mierla}, {Pomoell},
  \& {Poedts}}]{2019A&A...626A.122S}
{Scolini}, C., {Rodriguez}, L., {Mierla}, M., {Pomoell}, J., \& {Poedts}, S.
  2019, \aap, 626, A122

\bibitem[{{Song} {et~al.}(2013){Song}, {Chen}, {Ye}, {Han}, {Du}, {Li},
  {Zhang}, \& {Hu}}]{2013ApJ...773..129S}
{Song}, H.~Q., {Chen}, Y., {Ye}, D.~D., {et~al.} 2013, \apj, 773, 129

\bibitem[{{Subramanian} {et~al.}(2012){Subramanian}, {Lara}, \&
  {Borgazzi}}]{2012GeoRL..3919107S}
{Subramanian}, P., {Lara}, A., \& {Borgazzi}, A. 2012, \grl, 39, L19107

\bibitem[{{Temmer} {et~al.}(2010){Temmer}, {Veronig}, {Kontar}, {Krucker}, \&
  {Vr{\v{s}}nak}}]{2010ApJ...712.1410T}
{Temmer}, M., {Veronig}, A.~M., {Kontar}, E.~P., {Krucker}, S., \&
  {Vr{\v{s}}nak}, B. 2010, \apj, 712, 1410

\bibitem[{{Temmer} {et~al.}(2008){Temmer}, {Veronig}, {Vr{\v{s}}nak},
  {Ryb{\'a}k}, {G{\"o}m{\"o}ry}, {Stoiser}, \&
  {Mari{\v{c}}i{\'c}}}]{2008ApJ...673L..95T}
{Temmer}, M., {Veronig}, A.~M., {Vr{\v{s}}nak}, B., {et~al.} 2008, \apjl, 673,
  L95

\bibitem[{{Thernisien} {et~al.}(2009){Thernisien}, {Vourlidas}, \&
  {Howard}}]{2009SoPh..256..111T}
{Thernisien}, A., {Vourlidas}, A., \& {Howard}, R.~A. 2009, \solphys, 256, 111

\bibitem[{{Thernisien} {et~al.}(2006){Thernisien}, {Howard}, \&
  {Vourlidas}}]{2006ApJ...652..763T}
{Thernisien}, A.~F.~R., {Howard}, R.~A., \& {Vourlidas}, A. 2006, \apj, 652,
  763

\bibitem[{{Veronig} {et~al.}(2018){Veronig}, {Podladchikova}, {Dissauer},
  {Temmer}, {Seaton}, {Long}, {Guo}, {Vr{\v{s}}nak}, {Harra}, \&
  {Kliem}}]{2018ApJ...868..107V}
{Veronig}, A.~M., {Podladchikova}, T., {Dissauer}, K., {et~al.} 2018, \apj,
  868, 107

\bibitem[{{Vourlidas} {et~al.}(2000){Vourlidas}, {Subramanian}, {Dere}, \&
  {Howard}}]{2000ApJ...534..456V}
{Vourlidas}, A., {Subramanian}, P., {Dere}, K.~P., \& {Howard}, R.~A. 2000,
  \apj, 534, 456

\bibitem[{{Vourlidas} {et~al.}(2003){Vourlidas}, {Wu}, {Wang}, {Subramanian},
  \& {Howard}}]{2003ApJ...598.1392V}
{Vourlidas}, A., {Wu}, S.~T., {Wang}, A.~H., {Subramanian}, P., \& {Howard},
  R.~A. 2003, \apj, 598, 1392

\bibitem[{{Vr{\v{s}}nak}(2001)}]{2001SoPh..202..173V}
{Vr{\v{s}}nak}, B. 2001, \solphys, 202, 173

\bibitem[{{Vr{\v{s}}nak}(2008)}]{2008AnGeo..26.3089V}
---. 2008, Annales Geophysicae, 26, 3089

\bibitem[{{Vr{\v{s}}nak} {et~al.}(2004{\natexlab{a}}){Vr{\v{s}}nak},
  {Mari{\v{c}}i{\'c}}, {Stanger}, \& {Veronig}}]{2004SoPh..225..355V}
{Vr{\v{s}}nak}, B., {Mari{\v{c}}i{\'c}}, D., {Stanger}, A.~L., \& {Veronig}, A.
  2004{\natexlab{a}}, \solphys, 225, 355

\bibitem[{{Vr{\v{s}}nak} {et~al.}(2007){Vr{\v{s}}nak}, {Mari{\v{c}}i{\'c}},
  {Stanger}, {Veronig}, {Temmer}, \& {Ro{\v{s}}a}}]{2007SoPh..241...85V}
{Vr{\v{s}}nak}, B., {Mari{\v{c}}i{\'c}}, D., {Stanger}, A.~L., {et~al.} 2007,
  \solphys, 241, 85

\bibitem[{{Vr{\v{s}}nak} {et~al.}(2004{\natexlab{b}}){Vr{\v{s}}nak},
  {Ru{\v{z}}djak}, {Sudar}, \& {Gopalswamy}}]{2004A&A...423..717V}
{Vr{\v{s}}nak}, B., {Ru{\v{z}}djak}, D., {Sudar}, D., \& {Gopalswamy}, N.
  2004{\natexlab{b}}, \aap, 423, 717

\bibitem[{{Vr{\v{s}}nak} \& {{\v{Z}}ic}(2007)}]{2007A&A...472..937V}
{Vr{\v{s}}nak}, B., \& {{\v{Z}}ic}, T. 2007, \aap, 472, 937

\bibitem[{{Zhang} {et~al.}(2001){Zhang}, {Dere}, {Howard}, {Kundu}, \&
  {White}}]{2001ApJ...559..452Z}
{Zhang}, J., {Dere}, K.~P., {Howard}, R.~A., {Kundu}, M.~R., \& {White}, S.~M.
  2001, \apj, 559, 452

\bibitem[{{Zhang} {et~al.}(2004){Zhang}, {Dere}, {Howard}, \&
  {Vourlidas}}]{2004ApJ...604..420Z}
{Zhang}, J., {Dere}, K.~P., {Howard}, R.~A., \& {Vourlidas}, A. 2004, \apj,
  604, 420

\bibitem[{{Zhu} {et~al.}(2020){Zhu}, {Qiu}, {Liewer}, {Vourlidas}, {Spiegel},
  \& {Hu}}]{2020ApJ...893..141Z}
{Zhu}, C., {Qiu}, J., {Liewer}, P., {et~al.} 2020, \apj, 893, 141

\bibitem[{{Zhuang} {et~al.}(2020){Zhuang}, {Lugaz}, {Gou}, {Ding}, \&
  {Wang}}]{2020ApJ...901...45Z}
{Zhuang}, B., {Lugaz}, N., {Gou}, T., {Ding}, L., \& {Wang}, Y. 2020, \apj,
  901, 45

\bibitem[{{Zhuang} {et~al.}(2018){Zhuang}, {Wang}, {Shen}, \&
  {Liu}}]{2018E&PP....2..112Z}
{Zhuang}, B., {Wang}, Y., {Shen}, C., \& {Liu}, R. 2018, Earth and Planetary
  Physics, 2, 112

\end{thebibliography}
	
\end{document}